\documentclass[twocolumn,english,aps,prl,twocolumn,superscriptaddress,showpacs,floatfix]{revtex4-1}
\usepackage{mathptmx}

\usepackage[T1]{fontenc}
\usepackage[latin9]{inputenc}
\usepackage{multirow}
\usepackage{amsmath}
\usepackage{amssymb}
\usepackage{graphicx}

\makeatletter

\providecommand{\tabularnewline}{\\}

\makeatother

\usepackage{babel}
\begin{document}

\title{Cubic Double Perovskites Host Noncoplanar Spin Textures}
\author{Joseph A. M. Paddison}
\email{paddisonja@ornl.gov}

\affiliation{Materials Science and Technology Division, Oak Ridge National Laboratory,
Oak Ridge, TN 37831, USA}
\author{Hao Zhang}
\affiliation{Department of Physics and Astronomy, University of Tennessee, Knoxville,
Tennessee 37996, USA}
\author{Jiaqiang Yan}
\affiliation{Materials Science and Technology Division, Oak Ridge National Laboratory,
Oak Ridge, TN 37831, USA}
\author{Matthew J. Cliffe}
\affiliation{School of Chemistry, University of Nottingham, Nottingham NG7 2RD,
UK}
\author{Seung-Hwan Do}
\affiliation{Materials Science and Technology Division, Oak Ridge National Laboratory,
Oak Ridge, TN 37831, USA}
\author{Shang Gao}
\affiliation{Materials Science and Technology Division, Oak Ridge National Laboratory,
Oak Ridge, TN 37831, USA}
\affiliation{Neutron Scattering Division, Oak Ridge National Laboratory, Oak Ridge,
Tennessee 37831, USA}
\author{Matthew B. Stone}
\affiliation{Neutron Scattering Division, Oak Ridge National Laboratory, Oak Ridge,
Tennessee 37831, USA}
\author{David Dahlbom}
\affiliation{Department of Physics and Astronomy, University of Tennessee, Knoxville,
Tennessee 37996, USA}
\author{Kipton Barros}
\affiliation{Theoretical Division, Los Alamos National Laboratory, Los Alamos,
New Mexico 87545, USA}
\author{Cristian D. Batista}
\affiliation{Department of Physics and Astronomy, The University of Tennessee,
Knoxville, Tennessee 37996, USA}
\author{Andrew D. Christianson}
\email{christiansad@ornl.gov}

\affiliation{Materials Science and Technology Division, Oak Ridge National Laboratory,
Oak Ridge, TN 37831, USA}
\begin{abstract}
Magnetic materials with noncoplanar magnetic structures can show unusual
physical properties driven by nontrivial topology. Topologically-active
states are often multi-$\mathbf{q}$ structures, which are challenging
to stabilize in models and to identify in materials. Here, we use
inelastic neutron-scattering experiments to show that the insulating
double perovskites Ba$_{2}$YRuO$_{6}$ and Ba$_{2}$LuRuO$_{6}$
host a noncoplanar 3-$\mathbf{q}$ structure on the face-centered
cubic lattice. Quantitative analysis of our neutron-scattering data
reveals that these 3-$\mathbf{q}$ states are stabilized by biquadratic
interactions. Our study identifies double perovskites as a highly
promising class of materials to realize topological magnetism, elucidates
the stabilization mechanism of the 3-$\mathbf{q}$ state in these
materials, and establishes neutron spectroscopy on powder samples
as a valuable technique to distinguish multi-$\mathbf{q}$ from single-$\mathbf{q}$
states, facilitating the discovery of topologically-nontrivial magnetic
materials.
\end{abstract}
\maketitle
Most magnetic materials order with simple magnetic structures in which
spins are collinear or coplanar. Noncoplanar magnetic structures are
relatively rare, but are of great current interest, because they can
exhibit topological character and exotic physical properties \citep{Tokura_2021,Shindou_2001}.
For example, the finite scalar spin chirality of noncoplanar spin
textures can generate a topological magneto-optical effect \citep{Feng_2020}
and anomalous quantum Hall effect \citep{Surgers_2014,Zhou_2016},
even in the absence of spin-orbit coupling. Topologically-nontrivial
spin textures are typically multi-$\mathbf{q}$ structures, which
superpose magnetic modulations with symmetry-related wavevectors $\mathbf{q}$
\citep{Shindou_2001}. Multi-$\mathbf{q}$ spin textures with long-wavelength
modulations, such as skyrmion and hedgehog crystals, are well-studied
as hosts of topology-driven phenomena \citep{Kurumaji_2019,Hirschberger_2019,Hirschberger_2020}.
In this context, multi-$\mathbf{q}$ antiferromagnets are increasingly
important \citep{Gao_2020}, because they offer higher densities of
topological objects with the potential to generate stronger physical
responses \citep{Gomonay_2018}.

\begin{figure*}
\includegraphics{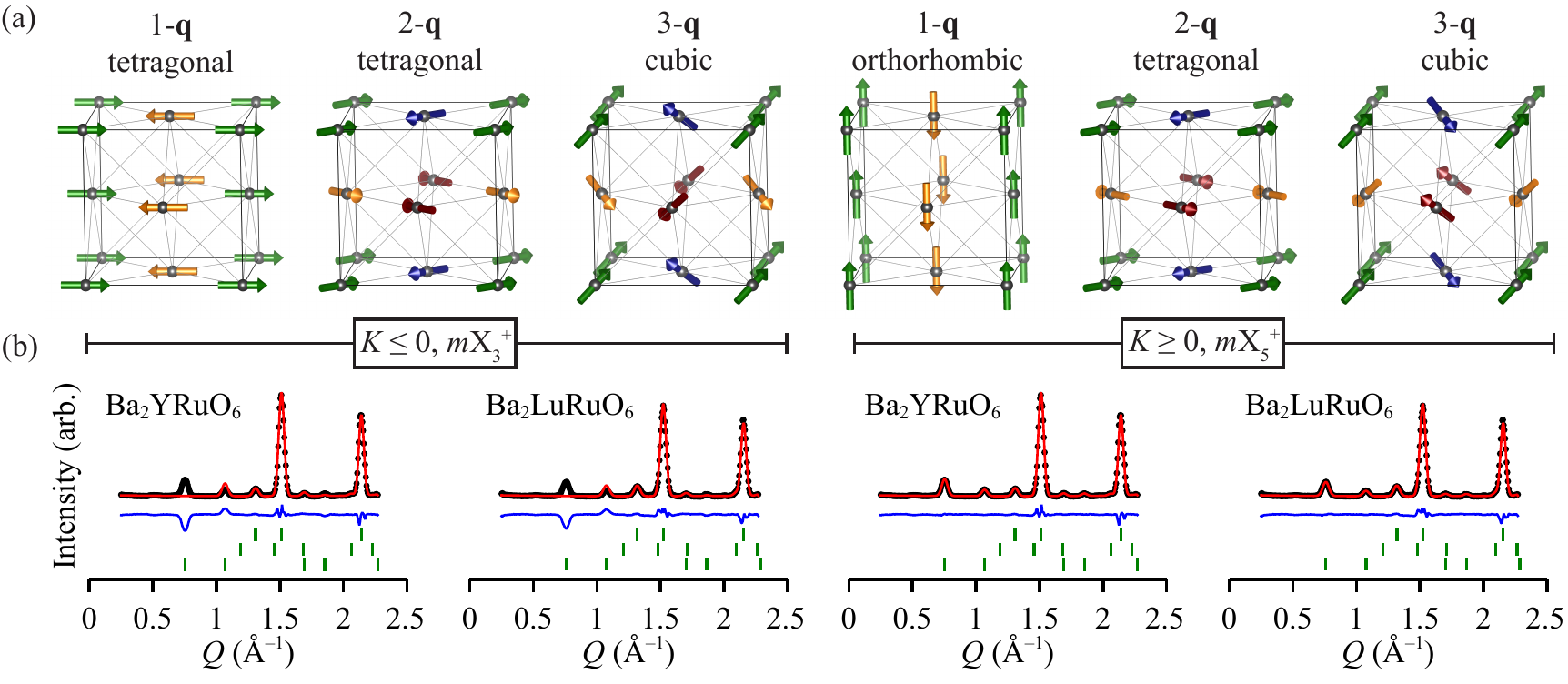}
\centering{}\caption{\label{fig:fig1} (a) Symmetry-allowed magnetic structures with propagation
vector $\mathbf{q}=[1,0,0]$ on the FCC lattice for Ba$_{2}$MRuO$_{6}$
(space group $Fm\bar{3}m$; $a=8.29$ and $8.24$\,\AA~for
$\textrm{M}=\textrm{Y}$ and Lu, respectively). The 1-$\mathbf{q}$,
2-$\mathbf{q}$, and 3-$\mathbf{q}$ structures are shown for the
$m\mathrm{X}_{3}^{+}$ irrep (left) and the $m\mathrm{X}_{5}^{+}$
irrep (right). Spins along different directions are colored differently;
note that 1-$\mathbf{q}$, 2-$\mathbf{q}$, and 3-$\mathbf{q}$ structures
have $[100]$, $\left\langle 110\right\rangle $, and $\left\langle 111\right\rangle $
spin directions, respectively. (b) Elastic scattering data ($-1.3\protect\leq E\protect\leq1.3$\,meV)
measured at $T=5$\,K with $E_{i}=11.8$\,meV for Ba$_{2}$YRuO$_{6}$
and Ba$_{2}$LuRuO$_{6}$ (black circles), Rietveld refinements (red
lines), and data -- fit (blue lines). Tick marks show (top to bottom):
nuclear, impurity M$_{2}$O$_{3}$, and magnetic phases. The $m\mathrm{X}_{3}^{+}$
irrep (left) does not reproduce our data, whereas the $m\mathrm{X}_{5}^{+}$
irrep (right) agrees well with our data. }
\end{figure*}

To probe the relationships between spin structure, interactions, topology,
and physical response, it is crucial to identify real materials that
host noncoplanar spin textures. This has proved a challenging task,
for three main reasons. First, it is necessary to identify noncoplanar
spin textures that are robust to subleading effects such as magnetic
anisotropies, spin-lattice coupling \citep{Penc_2004,Wang_2008},
fluctuations \citep{Gvozdikova_2005,Schick_2020,Singh_2003,McClarty_2014},
and anisotropic interactions \citep{Maksimov_2019}, which usually
favor collinear states. Second, most noncoplanar states are found
in metals, such as USb \citep{Jensen_1981,Halg_1986} and $\gamma$-Mn
alloys \citep{Hirai_1985,Kawarazaki_1988,Kawarazaki_1990,Long_1990,Fishman_2000,Hanke_2017},
and are often stable only under an applied magnetic field \citep{Kurumaji_2019,Khanh_2020}.
On the one hand, itinerant electrons can support the generation of
physical responses; on the other hand, modeling the magnetic interactions
of metals presents fundamental challenges \citep{Agterberg_2000,Hayami_2014,Jo_1983,Matsuura_2009,Hayami_2021,Hayami_2021c},
such that insulators are often more suitable as model materials. Third,
powder neutron-diffraction measurements play a central role in solving
magnetic structures, but suffer from a ``multi-$\mathbf{q}$ problem'':
Such measurements are generally unable to distinguish 1-$\mathbf{q}$
from multi-$\mathbf{q}$ structures \citep{Kouvel_1963}. Therefore,
multi-$\mathbf{q}$ spin textures are challenging to stabilize in
models, and to identify in real materials.

Here, we identify the Mott-insulating double perovskites
Ba$_{2}$YRuO$_{6}$ and Ba$_{2}$LuRuO$_{6}$ \citep{Battle_1989,Carlo_2013,Aharen_2009,Nilsen_2015}
as prototypical examples of noncoplanar 3-$\mathbf{q}$ magnetism
on the face-centered cubic (FCC) lattice in zero magnetic field. We
obtain evidence for 3-$\mathbf{q}$ magnetism from a spin-wave analysis
of neutron spectroscopy data. By optimizing the magnetic structure
and interactions simultaneously against our data, we show that the
3-$\mathbf{q}$ structure is stabilized by biquadratic interactions
within an antiferromagnetic Heisenberg-Kitaev model. Our study experimentally
establishes that noncoplanar multi-$\mathbf{q}$ states are stabilized
in frustrated FCC antiferromagnets, identifies cubic double perovskites
as model materials to realize this behavior, and identifies guiding
principles to facilitate design of materials with noncoplanar
states.

Our study is motivated by theoretical results for the FCC antiferromagnet
\citep{Gvozdikova_2005,Cook_2015,Balla_2020,Diop_2022}. The nearest-neighbor
Heisenberg-Kitaev spin Hamiltonian on the FCC lattice can be written
as
\begin{align}
H & =J\sum_{\left\langle i,j\right\rangle }\mathbf{S}_{i}\cdot\mathbf{S}_{j}+K\sum_{\left\langle i,j\right\rangle _{\gamma}}S_{i}^{\gamma}S_{j}^{\gamma},\label{eq:JK_model}
\end{align}
where $\mathbf{S}_{i}$ is a Ru$^{5+}$ spin with quantum number $S=3/2$,
$J$ and $K$ denote the Heisenberg and Kitaev interactions, respectively,
and $\gamma\in\left\{ x,y,z\right\} $ is perpendicular to the cubic
plane containing the bond between neighbors $\left\langle i,j\right\rangle $.
For antiferromagnetic $J>0$ only, the model is frustrated, and orderings
with $\mathbf{q}\in[1,q,0]$ are degenerate \citep{Gvozdikova_2005,Balla_2020,Diop_2022}.
The degenerate manifold includes $\mathbf{q}=[1,0,0]$ (``Type I'')
ordering, which is favored by fluctuations \citep{Gvozdikova_2005,Schick_2020,Schick_2022}
and is observed in Ba$_{2}$YRuO$_{6}$ and Ba$_{2}$LuRuO$_{6}$
\citep{Battle_1989}. Henceforth, we therefore restrict our discussion
to $\mathbf{q}=[1,0,0]$ ordering. For a collinear structure, spins
may be either parallel or perpendicular to $\mathbf{q}$; the former is favored by $K<0$ and the latter by $K>0$ \citep{Cook_2015,Balla_2020,Diop_2022}. 

Figure~\ref{fig:fig1}(a) shows the collinear (1-$\mathbf{q}$) and
noncollinear (multi-$\mathbf{q}$) structures associated with Type
I antiferromagnetism. A remarkable property of the FCC lattice is
that 1-$\mathbf{q}$, 2-$\mathbf{q}$, and 3-$\mathbf{q}$ structures
are energetically degenerate for \emph{all} bilinear interactions
for which Type I ordering is stable \citep{Balla_2020,Diop_2022}.
Moreover, uniaxial anisotropy ($\sim$$S_{z}^{2}$) and antisymmetric
exchange terms are forbidden by $Fm\bar{3}m$ symmetries, and quartic
anisotropy ($\sim$$S_{x}^{4}+S_{y}^{4}+S_{z}^{4}$) is forbidden
for $S=3/2$ operators in a cubic environment. Consequently, interactions
that would usually favor collinear magnetic structures are inactive
in the $S=3/2$ FCC antiferromagnet. This remarkable property potentially
allows noncollinear structures to appear.

To identify candidate systems for 3-$\mathbf{q}$ spin textures among
the diverse magnetic ground states of double perovskites \citep{Gangopadhyay_2016,Paramekanti_2018,Bos_2004,Taylor_2016,Taylor_2018,Gao_2020a,Paramekanti_2020,Maharaj_2020,Iwahara_2018},
we consider two criteria: Type I antiferromagnetic ordering, and
strictly cubic symmetry below the magnetic ordering temperature, $T_{\mathrm{N}}$.
The second criterion is key because 3-$\mathbf{q}$ structures
have cubic symmetry, while 1-$\mathbf{q}$ and 2-$\mathbf{q}$ structures
have tetragonal or orthorhombic symmetry 
that could drive a crystallographic distortion \emph{via }spin-lattice
coupling {[}Figure~\ref{fig:fig1}(a){]}. We investigate Ba$_{2}$YRuO$_{6}$ and Ba$_{2}$LuRuO$_{6}$
because they are chemically well-ordered and show no evidence for
low-temperature deviations from cubic symmetry \citep{Battle_1989,Aharen_2009}.
Moreover, recent first-principles calculations predict that their
magnetic structures might not be collinear \citep{Fang_2019}, in
apparent contradiction with interpretations of previous experiments
\citep{Battle_1989}.

We prepared $\sim$$8$\,g polycrystalline samples of Ba$_{2}$YRuO$_{6}$
and Ba$_{2}$LuRuO$_{6}$ by solid-state reaction \citep{SI}. Rietveld
refinement revealed stoichiometric samples with minor Lu$_{2}$O$_{3}$ ($1.94$\,wt.\%) or Y$_{2}$O$_{3}$ ($0.65$\,wt.\%) impurities.
The magnetic ordering temperature $T_{\mathrm{N}}\approx37$\,K is
the same for both samples, and is suppressed compared to the Weiss
temperature $\theta\sim-500$\,K, indicating strong magnetic frustration
\citep{Aharen_2009}. We performed inelastic neutron-scattering measurements
on the SEQUOIA instrument at ORNL \citep{Granroth_2010} using incident
neutron energies $E_{i}=62$ and $11.8$\,meV, yielding elastic energy
resolutions $\delta\mathrm{_{ins}}\approx1.68$ and $0.27$\,meV,
respectively.

Figure~\ref{fig:fig1}(b) shows magnetic Rietveld refinements to
our elastic neutron-scattering data, measured with $E_{i}=11.8$\,meV
at $T\approx5$\,K. Applying the $\mathbf{q}=[1,0,0]$ propagation
vector to $Fm\bar{3}m$ crystal symmetry generates two magnetic irreducible
representations (irreps), notated $m\mathrm{X}_{3}^{+}$ and $m\mathrm{X}_{5}^{+}$
\citep{Cracknell_1979,Wills_2001}. These irreps can be distinguished
by their magnetic Bragg profiles. The $m\mathrm{X}_{5}^{+}$ irrep
agrees well with our elastic-scattering data for both materials; we
obtain ordered magnetic moment lengths of $2.56(2)$ and $2.43(2)$\,$\mu_{\mathrm{B}}$
per Ru for Ba$_{2}$YRuO$_{6}$ and Ba$_{2}$LuRuO$_{6}$, respectively,
from Rietveld refinement. Since the magnetic form factor for Ru$^{5+}$
is not known, we tested several $4d$ magnetic form factors \citep{Brown_2004};
while this choice does not qualitatively affect our results, the form
factor for Zr$^{+}$ (isoelectronic with Ru$^{5+}$) yields optimal
agreement with our data and is used throughout. In contrast to the
$m\mathrm{X}_{5}^{+}$ irrep, the $m\mathrm{X}_{3}^{+}$ irrep strongly
disagrees with our data, as it yields zero intensity for the strong
$(100)$ magnetic Bragg peak. This can be understood intuitively for
a collinear 1-$\mathbf{q}$ structure, because neutrons are only sensitive
to spin components perpendicular to the scattering wavevector, and
the $m\mathrm{X}_{3+}$ irrep has $\mathbf{S}\parallel\mathbf{q}$
while the $m\mathrm{X}_{5+}$ irrep has $\mathbf{S}\perp\mathbf{q}$
{[}Figure~\ref{fig:fig1}(a){]}. A previous elastic neutron-scattering
study of Ba$_{2}$YRuO$_{6}$ and Ba$_{2}$LuRuO$_{6}$ considered
only collinear 1-$\mathbf{q}$ structures \citep{Battle_1989}, but
could not rule out multi-$\mathbf{q}$ structures, due to the multi-$\mathbf{q}$
problem.

\begin{figure}
\begin{centering}
\includegraphics{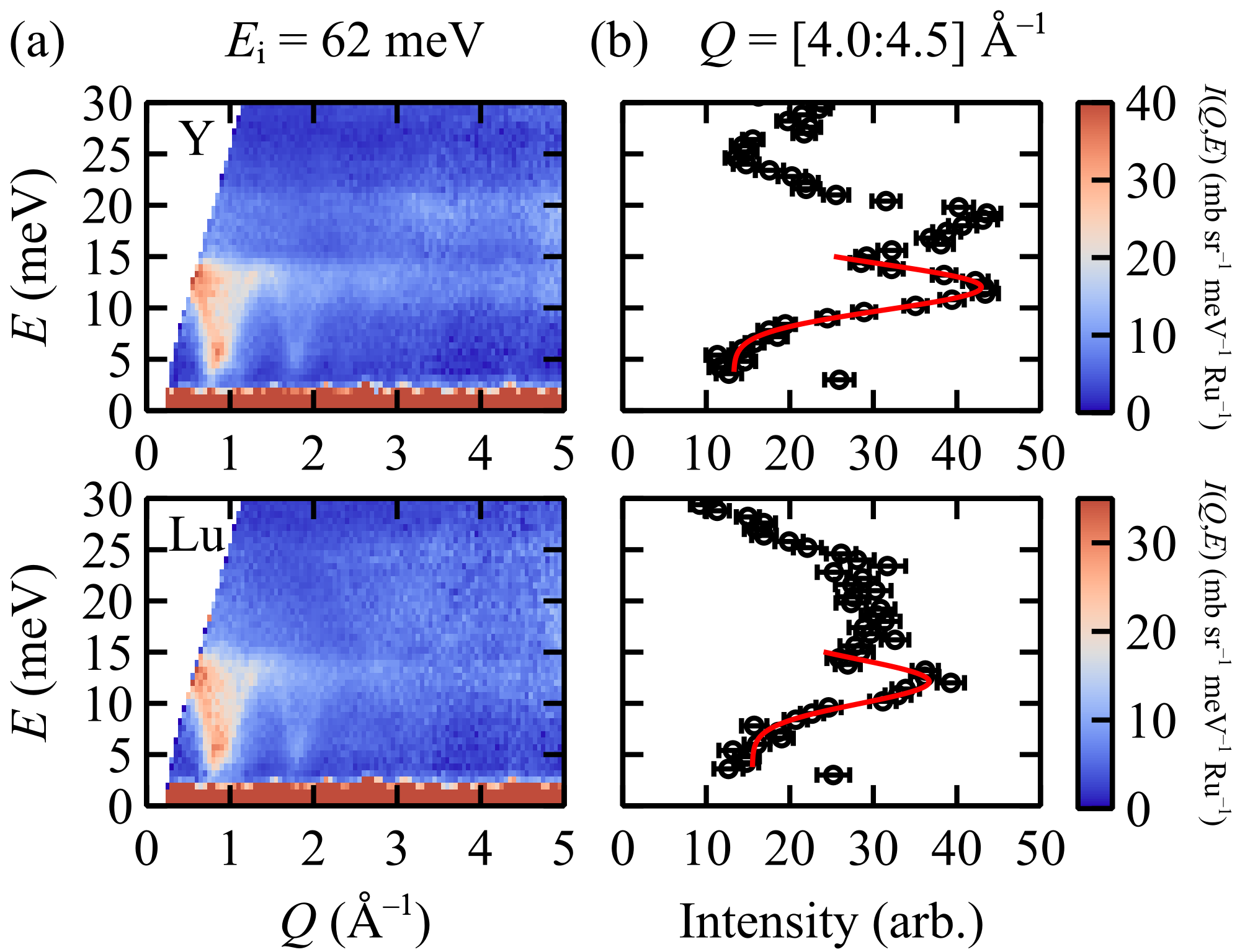}
\par\end{centering}
\centering{}\caption{\label{fig:fig2}Broadband inelastic neutron-scattering data ($E_{i}=62$\,meV)
measured at $T=5$\,K for Ba$_{2}$YRuO$_{6}$ (upper panels) and
Ba$_{2}$LuRuO$_{6}$ (lower panels), showing (a) intensity as a color
plot, and (b) energy dependence integrated over $4.0 \leq Q \leq4.5$\,\AA$^{-1}$,
where experimental data are shown as black circles, and Gaussian fits
to the $\sim$$14$\,meV phonon band as red lines.}
\end{figure}

To overcome the multi-\textbf{q} problem, we consider the energy dependence
of our neutron-scattering data \citep{Paddison_2021}. Figure~\ref{fig:fig2}(a)
shows our inelastic data measured with $E_{i}=62$\,meV at $T\approx5$\,K.
A structured inelastic signal appears at $T<T_{\mathrm{N}}$ for small
scattering wavevectors, $Q\lesssim 2$\,\AA$^{-1}$,
which we identify as magnon scattering. The top of the magnetic band
overlaps with an intense phonon signal for $Q\gtrsim 2$\,\AA$^{-1}$.
Figure~\ref{fig:fig2}(b) shows the scattering intensity integrated
over $4.0\leq Q\leq4.5$\,\AA$^{-1}$, from
which we extract the average energy $E_{\mathrm{ph}}$ and width $\sigma_{\mathrm{ph}}$
of this phonon band \emph{via }Gaussian fits for each material. The
energy overlap of magnon and phonon modes suggests that spin-lattice
coupling may be significant, which we consider further below.

\begin{figure}
\begin{centering}
\includegraphics{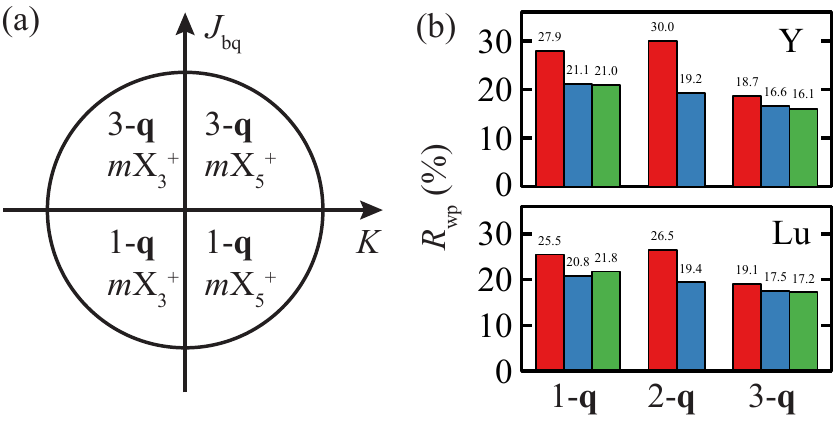}
\par\end{centering}
\centering{}\caption{\label{fig:fig3}(a) Schematic phase diagram showing the magnetic
ground states of the $J$-$K$-$J_{\mathrm{bq}}$ model. (b) Goodness-of-fit
metric $R_{\mathrm{wp}}$ for candidate magnetic structures and interaction
models of Ba$_{2}$YRuO$_{6}$ (upper graph) and Ba$_{2}$LuRuO$_{6}$
(lower graph). The graphs show $R_{\mathrm{wp}}$ for refinements
of the Heisenberg-Kitaev ($J$-$K$) model including a third refined
parameter $\Gamma$ (red bars), $J_{2}$ (blue bars), or $J_{\mathrm{bq}}$
(green bars); note that the 2-$\mathbf{q}$ structure is stable only
for $J_{\mathrm{bq}}=0$.}
\end{figure}

Our starting point for modeling the magnetic scattering
is the Heisenberg-Kitaev model, Eq.~(\ref{eq:JK_model}). For all
models, we require $J>0$ and $K>0$ to stabilize $m\mathrm{X}_{5}^{+}$
ordering. We consider three additional interactions in turn. First,
the symmetric off-diagonal interaction $H_{\Gamma}=\Gamma\sum_{\left\langle i,j\right\rangle _{\gamma}}\left(S_{i}^{\alpha}S_{j}^{\beta}+S_{i}^{\beta}S_{j}^{\alpha}\right)$
is the only additional bilinear nearest-neighbor interaction allowed
by symmetry. Second, the Heisenberg next-nearest neighbor interaction
$H_{2}=J_{2}\sum_{\left\langle \left\langle i,j\right\rangle \right\rangle }\mathbf{S}_{i}\cdot\mathbf{S}_{j}$
has been invoked for Ba$_{2}$YRuO$_{6}$ \citep{Nilsen_2015}; we
require $J_{2}\leq0$ to stabilize Type I ordering. Third, the nearest-neighbor
biquadratic coupling $H_{\mathrm{bq}}=J_{\mathrm{bq}}\sum_{\left\langle i,j\right\rangle }(\mathbf{S}_{i}\cdot\mathbf{S}_{j})^{2}$
has been invoked in density-functional-theory calculations for 4$d$
double perovskites due to their increased electron hopping relative
to $3d$ analogs \citep{Fang_2019}. For $J_{\mathrm{bq}}=0$, the
classical energy of 1-$\mathbf{q}$, 2-$\mathbf{q}$, and 3-$\mathbf{q}$
structures is equal for all $K$, $\Gamma$, and $J_{2}$ that stabilize
Type I ordering. Nonzero $J_{\mathrm{bq}}$ removes this degeneracy,
and stabilizes 1-$\mathbf{q}$ ordering for $J_{\mathrm{bq}}<0$ and
3-$\mathbf{q}$ ordering for $J_{\mathrm{bq}}>0$ {[}Figure~\ref{fig:fig3}(a){]}.
Importantly, since single-ion anisotropies are forbidden for $S=3/2$
in a cubic environment, biquadratic exchange is the only physically-plausible
mechanism that can remove the degeneracy of 1-$\mathbf{q}$ and 3-$\mathbf{q}$
structures.

\begin{figure*}
\begin{centering}
\includegraphics{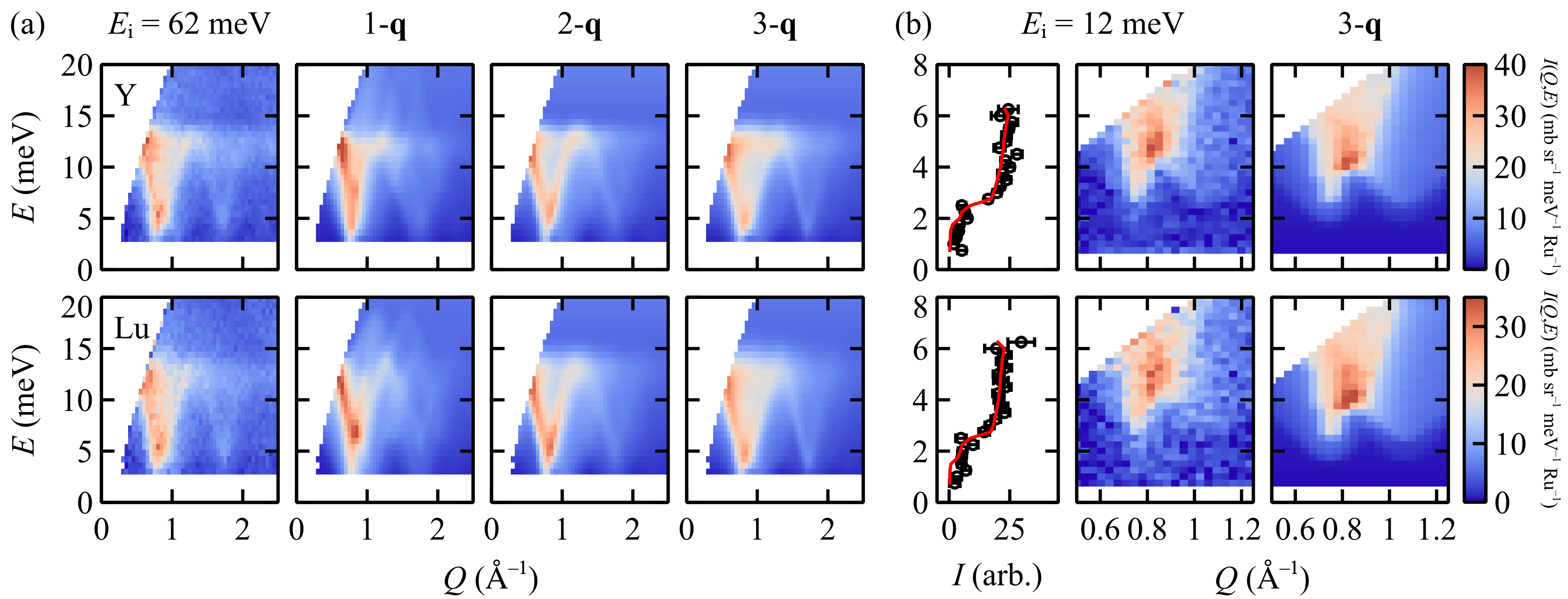}
\par\end{centering}
\centering{}\caption{\label{fig:fig4}(a) Broadband inelastic neutron-scattering data ($E_{i}=62$\,meV)
and optimal spin-wave fits for different magnetic structures, showing
(left to right) experimental data, 1-$\mathbf{q}$ fit, 2-$\mathbf{q}$
fit, and 3-$\mathbf{q}$ fit. (b) Low-energy inelastic neutron-scattering
data ($E_{i}=11.8$\,meV) and $3$-$\mathbf{q}$ model calculations,
showing (left to right) a cut at $Q=0.7450\pm0.0175$\,\AA$^{-1}$
comparing experimental data (black circles) and spin-wave fit (red
lines), experimental data as a $Q$-$E$ slice, and spin-wave calculation.}
\end{figure*}

We performed extensive fits to our inelastic neutron-scattering data
to optimize the magnetic structure and interactions simultaneously.
For each structure associated with the $m\mathrm{X}_{5}^{+}$ irrep
(1-$\mathbf{q}$, 2-$\mathbf{q}$, or 3-$\mathbf{q}$), we optimized
three spin Hamiltonian parameters ($J$, $K$, and either $\Gamma$,
$J_{2}$, or $J_{\mathrm{bq}}$) against the broadband inelastic data
shown in Figure~\ref{fig:fig4}(a) and the energy dependence near
the $(100)$ magnetic Bragg position shown in Figure~\ref{fig:fig4}(b).
The powder-averaged magnon spectrum was calculated within a renormalized
linear spin-wave theory \citep{Ader_2001} using the SpinW program
\citep{Toth_2015}. The renormalization factor, which takes into account
higher-order corrections in the $1/S$ expansion, is strictly necessary
to extract a correct value of $J_{\mathrm{bq}}$, since the unrenormalized
spin-wave theory would lead to a value of $J_{\mathrm{bq}}$ that
is 2.25 times smaller than the correct value \citep{Dahlbom_2022}. The parameter values were optimized
to minimize the sum of squared residuals using nonlinear least-squares
refinement \citep{SI}. We calculated the energy-dependent broadening
of the magnon spectrum as $\delta(E)=\mathrm{\delta}_{\mathrm{ins}}(E)+Ae^{-(E-E_{\mathrm{ph}})^{2}/2\delta_{\mathrm{ph}}^{2}}$,
where $\delta(E)$ is the overall Gaussian energy width, $\delta_{\mathrm{ins}}(E)$
is the instrumental resolution, and $A$ is a refined parameter that
phenomenologically accounts for magnon broadening due to coupling
with phonons at $E\sim E_{\mathrm{ph}}$. 

\begin{table}
\begin{centering}
\begin{tabular}{c|cccc}
 & $J$ (K) & $K$ (K) & $J_{\mathrm{bq}}$ (K) & $A_{\mathrm{}}$ (meV)\tabularnewline
\hline 
Ba$_{2}$YRuO$_{6}$ & $21.85(3)$ & $0.39(1)$ & $1.32(2)$ & $0.97(3)$\tabularnewline
\multirow{1}{*}{Ba$_{2}$LuRuO$_{6}$} & $22.27(4)$ & $0.36(2)$ & $1.17(3)$ & $2.25(5)$\tabularnewline
\hline 
\end{tabular}
\par\end{centering}
\caption{\label{tab:params}Refined values of magnetic interaction parameters
for the $J$-$K$-$J_{\mathrm{bq}}$ model and 3-$\mathbf{q}$ structure.
Uncertainties indicate $1\sigma$ statistical confidence intervals.}
\end{table}

Figure~\ref{fig:fig4}(a) compares our broadband inelastic data ($E_{i}=62$\,meV)
with the best fit for each of the 1-$\mathbf{q}$, 2-$\mathbf{q}$,
and 3-$\mathbf{q}$ structures. The data show two V-shaped features
centered at $\approx0.85$ and $\approx1.70$\,\AA$^{-1}$,
with a sharp cutoff of magnetic signal for energies above $\sim$$14$\,meV.
For both materials, these characteristics are best reproduced by the
3-$\mathbf{q}$ structure, while the 1-$\mathbf{q}$ structure disagrees
with our experimental data. These observations are confirmed by the
goodness-of-fit metric $R_{\mathrm{wp}}$, shown in Figure~\ref{fig:fig3}(b).
For both materials and for every interaction model we considered,
the 3-$\mathbf{q}$ structure yields better agreement with our data
than the 1-$\mathbf{q}$ or 2-$\mathbf{q}$ structures. Notably, the
goodness-of-fit is more sensitive to the structure than the precise magnetic interactions; indeed, the main differences between
1-$\mathbf{q}$ and 3-$\mathbf{q}$ spectra are apparent for Heisenberg
exchange only \citep{SI}. The global best fit is for the 3-$\mathbf{q}$
structure and $J$, $K$, and $J_{\mathrm{bq}}$ interactions with
the refined values given in Table~\ref{tab:params}. The refined
values of $A$ indicate significant magnon broadening, which is larger
for Ba$_{2}$LuRuO$_{6}$ and is likely due to magnon-phonon coupling.
Importantly, for both materials, the biquadratic
term is significant, with $J_{\mathrm{bq}}/J\sim0.06$. Hence, our
key results are that only the 3-$\mathbf{q}$ spin texture agrees
well with our neutron data, and this state is stabilized by biquadratic
interactions in Ba$_{2}$YRuO$_{6}$ and Ba$_{2}$LuRuO$_{6}$.

Our model provides insight into the mechanism of gap opening \citep{Carlo_2013}.
Figure~\ref{fig:fig4}(b) compares our low-energy inelastic data
($E_{i}=11.8$\,meV) with the 3-$\mathbf{q}$ magnon spectrum for
the optimal $J$-$K$-$J_{\mathrm{bq}}$ model {[}Table~\ref{tab:params}{]}.
This calculation reproduces the observed $\approx2.8$\,meV gap,
unlike the $J$-$K$-$J_{2}$ model that yields the next-best
$R_{\mathrm{wp}}$ \citep{SI}. Since single-ion anisotropies are
forbidden here, the mechanism of gap opening is subtle. If $K=0$,
there is no gap, because the energy of the Heisenberg and biquadratic
terms is unchanged by global spin rotations. For $K>0$, whether a
gap opens depends on both structure and interactions. If the structure
is 1-$\mathbf{q}$ with $J_{\mathrm{bq}}<0$, the classical energy
is unchanged by global spin rotations in the plane perpendicular to
$\mathbf{q}$. In this case, there is no gap at the linear spin-wave
level; a gap is generated only by magnon interactions in the quantum
($S=1/2$) limit \citep{Aczel_2016}. By contrast, if the structure
is 3-$\mathbf{q}$ with $J_{\mathrm{bq}}>0$, a gap is present at
the linear spin-wave level, because $J_{\mathrm{bq}}>0$ and $K>0$
together favor $\left\langle 111\right\rangle $ spin alignment. Since
Ba$_{2}$YRuO$_{6}$ and Ba$_{2}$LuRuO$_{6}$ are not in the quantum
limit, the experimental observation of a gap supports the presence
of biquadratic and Kitaev interactions in a 3-$\mathbf{q}$ structure. 

We have shown that the magnetic ground states of Ba$_{2}$YRuO$_{6}$
and Ba$_{2}$LuRuO$_{6}$ are noncoplanar 3-$\mathbf{q}$ structures
stabilized by biquadratic interactions. Macroscopic topological physical
responses may be generated synthesizing thin films of these materials
with $[111]$ strain \citep{Wang_2007}. Our experimental results
strikingly confirm recent first-principles predictions \citep{Fang_2019}.
The positive sign of $J_{\mathrm{bq}}$ suggests that the effect of inter-site electron hopping outweighs spin-lattice coupling,
since the latter would give a negative contribution to $J_{\mathrm{bq}}$ \citep{Penc_2004,Wang_2008}.
Crucially, we quantify the magnetic interactions that stabilize the
noncoplanar state, in contrast to other proposed 3-$\mathbf{q}$ structures
in NiS$_{2}$ \citep{Kikuchi_1978,Yosida_1981,Higo_2015}, MnTe$_{2}$
\citep{Burlet_1997}, and UO$_{2}$ \citep{Frazer_1965,Faber_1976,Caciuffo_1999,Dudarev_2019},
where the relevant interactions are not yet well understood. Our work
provides several guiding principles to facilitate the identification
of multi-$\mathbf{q}$ spin textures. First, the near-degeneracy of
1-$\mathbf{q}$ and multi-$\mathbf{q}$ structures on the FCC lattice
makes double perovskites enticing systems. In candidate materials,
the crystal symmetry should be higher than a 1-$\mathbf{q}$ model
would imply. Second, magnets that are not deep inside the Mott-insulating
regime are expected to have larger $J_{\mathrm{bq}}$ and, consequently,
more robust 3-$\mathbf{q}$ orderings. This criterion hints that cubic
Ba$_{2}$YOsO$_{6}$ \citep{Kermarrec_2015,Maharaj_2018} may also
host a 3-$\mathbf{q}$ state, due to its extended Os $5d$ orbitals,
potentially offering a route to investigate the effect of increased
electron hopping. For small $J_{\mathrm{bq}}$, we anticipate a thermally-induced
transition from 3-$\mathbf{q}$ to 1-$\mathbf{q}$ ordering, since
thermal fluctuations favor collinear states. Third, quartic single-ion
anisotropy may play a role in FCC magnets with $S>3/2$; in particular,
easy-$\langle111\rangle$ axis anisotropy should favor 3-$\mathbf{q}$
ordering. Finally, our key methodological insight is that refining
the magnetic structure and interactions simultaneously enables 1-$\mathbf{q}$
and multi-$\mathbf{q}$ structures to be distinguished on the FCC
lattice, even when single-crystal samples are not available. 

\acknowledgements{This work was supported by the U.S. Department of Energy, Office of
Science, Basic Energy Sciences, Materials Sciences and Engineering
Division. This research used resources at the Spallation Neutron Source,
a DOE Office of Science User Facility operated by the Oak Ridge National
Laboratory.}


\end{document}